\documentclass[amssymb,prb,twocolumn,showpacs]{revtex4}
\usepackage{bm}
\usepackage{graphicx}
\usepackage{epsfig}
\usepackage{amsmath}
\newfont{\bl}{cmbxsl10 scaled\magstep1}
\begin{document}
\title{Spin-dependent tunneling in modulated structures of (Ga,Mn)As}
\author{P. Sankowski} \affiliation{Institute of Physics,
  Polish Academy of Sciences, al.~Lotnik\'ow 32/46, PL 02-668 Warszawa, Poland }
\author{P. Kacman} \affiliation{Institute of Physics,
  Polish Academy of Sciences, al.~Lotnik\'ow 32/46, PL 02-668 Warszawa, Poland }
\author{J. A. Majewski} \affiliation{Institute of Theoretical Physics and Interdisciplinary Center for Materials Modeling,
Warsaw University, ul. Ho\.za 69, PL 00-681 Warszawa, Poland}
\author{T. Dietl} \affiliation{Institute of Physics,
  Polish Academy of Sciences and ERATO Semiconductor Spintronics
  Project, al.~Lotnik\'ow 32/46, PL 02-668 Warszawa, Poland \\
Institute of Theoretical Physics, Warsaw University, ul. Ho\.za 69, 00-681 Warszawa, Poland}

\date{\today}
\begin{abstract}
A model of coherent tunneling, which combines multi-orbital
tight-binding approximation with Landauer-B\"uttiker formalism, is
developed and applied to all-semiconductor heterostructures
containing (Ga,Mn)As ferromagnetic layers. A comparison  of
theoretical predictions and experimental results on spin-dependent
Zener tunneling, tunneling magnetoresistance (TMR), and anisotropic
magnetoresistance (TAMR) is presented. The dependence of spin
current on carrier density, magnetization orientation, strain,
voltage bias, and spacer thickness is examined theoretically in
order to optimize device design and performance.
\end{abstract}
\maketitle

\section{Introduction}
Metallic magnetic tunnel junctions (MTJs) are important building
blocks of the already existing spintronic devices, such as magnetic
random access memories, magnetic heads, and sensors. The MTJ
structure consists of ferromagnetic layers separated by a thin
insulating barrier through which the carriers tunnel. The resistance
of such a junction depends on the relative alignment of the
magnetization vectors in the ferromagnetic layers, {\em i.~e.}, the
structure exhibits the tunnelling magnetoresistance  (TMR) effect.

All-semiconductor MTJ structures offer potential for precise control
of interfaces and barrier properties, particularly in the case of
III-V compounds, for which epitaxial growth of complex
heterostructures containing ferromagnetic (Ga,Mn)As or (In,Mn)As
layers is especialy advanced.\cite{Mats02} Extensive studies of MTJs
with (Ga,Mn)As ferromagnetic contacts carried out by various groups
resulted in  an increase of the observed TMR ratio from about 70\%
reported by Tanaka and Higo\cite{tanaka} to values higher than
250\%.\cite{mattana,chiba,elsen} Another key factor for developing
novel functional semiconductor spintronic devices is an efficient
electrical injection of spin polarized carriers. Here again the
III-V ferromagnetic p-type semiconductor (Ga,Mn)As with its high
spin polarization\cite{bib-7-dietl-01} appears as a promising
material. The electrical spin injection from p-(Ga,Mn)As into
non-magnetic semiconductor has first been achieved by injection of
spin polarized holes.\cite{ohno99} Later, injection of spin
polarized electrons was demonstrated employing interband tunneling
from the valence band of (Ga,Mn)As into the conduction band of an
adjacent n-GaAs in a Zener-Esaki diode.\cite{Kohd01John02} Recently,
a very high spin polarization of the injected electron current ({\it
ca} 80\%) was obtained in such devices.\cite{bib-4-van-dorpe,Kohd06}
It should be mentioned that both effects, the TMR and the spin
polarization of tunneling current in the Zener-Esaki diode, decrease
rapidly with the increase of the applied bias -- a phenomenon
observed also in the metallic TMR structures, and still not fully
understood. Finally, it seems that TMR is sensitive to the direction
of the applied magnetic field in respect to the direction of current
and crystallographic axes. This so-called tunnel anisotropic
magnetoresistance (TAMR) effect was observed in structures
containing a single ferromagnetic electrode\cite{gould,gryglas} as
well as in typical TMR MTJ with two ferromagnetic
contacts.\cite{ruster,giddings}

These challenging experimental findings call for a theory that would
describe the tunneling in semiconductor MTJs and would indicate the
ways for optimized design of the devices. Since the ferromagnetic
coupling in (Ga,Mn)As is mediated by the
holes,\cite{Mats02,bib-7-dietl-01} a meaningful theory has to take
into account the entire complexity of the valence band, including
the spin-orbit interaction. Furthermore, the intermixing of valence
bands caused by spin-orbit coupling shortens the spin diffusion
length and makes it comparable to the phase coherence length. This
renders the models based on the classical spin-diffusion equation,
which describe satisfactorily spin transport phenomena in metallic
MTJs, non applicable directly to the structures containing layers of
hole-controlled diluted ferromagnetic semiconductors.

The model of vertical transport in modulated structures of magnetic
semiconductors described here combines the two-terminal
Landauer-B\"uttiker formalism with the empirical multi-orbital
tight-binding description of the semiconductor band structure. In
this way, the quantum character of spin transport over the length
scale relevant for the devices in question is taken into account.
Furthermore, the tight-binding approach, in contrast to $kp$ models
employed so-far,\cite{Petu04Brey04} allows for a proper description
of effects crucial for spin transport in heterostructures such as
atomic structure of interfaces,  effects of Rashba and Dresselhaus
terms as well as tunneling involving ${\bm k}$ states away from the
center of the Brillouin zone. Our model has recently been applied to
describe selected features of Zener-Esaki
diodes\cite{van-dorpe,sankowski} and TMR devices\cite{sankowski} as
well as it was adopted to examine an intrinsic domain-wall
resistance in (Ga,Mn)As.\cite{Oszw06}

The remaining part of the present paper is organized as follows. In
Section II we present our model, specifying the tight-binding
parametrization, scattering formalism, and the way transfer
coefficients are determined. In Section III, the calculated
dependencies of spin polarization of the current in the Zener-Esaki
diode on carrier density, trigonal distortion, and magnetization
direction are shown and compared with available experimental
findings. The calculated TMR and TAMR ratios for structures
containing two ferromagnetic (Ga,Mn)As contacts, and their
dependencies on the voltage bias, trigonal deformation, and on the
width of non-magnetic spacer layer are presented and discussed in
Section IV in reference to experimental results. Section V contains
conclusions emerging from our work, particularly concerning possible
optimizations of device performance.

\section{Theoretical Model}
We consider a prototype heterostructure, which is uniform and
infinite in the $x$ and $y$ directions and has modulated
magnetization along the $z$ growth direction. The heterostructure is
connected to two semi-infinite bulk contacts denoted by $L$ and $R$
and biased. In all cases considered, spin polarized carriers are
injected from the ferromagnetic left lead. Our goal is to calculate
the electric current in the structure and the degree of current spin
polarization outside the left lead. Typical length of the studied
structures is comparable to the phase coherence length. Therefore,
we restrict ourselves to the vertical coherent transport regime that
we treat within Landauer-B\"uttiker formalism, where the current is
determined by the transmission probability from the ingoing Bloch
state at the left contact to the outgoing Bloch state at the right
contact. In the presence of spin-orbit coupling, the spin is not a
good quantum number. The only preserved quantities in tunneling are
the energy $E$ and, due to spatial in-plane symmetry of our
structures, the in-plane wave vector ${\bm k}_{\|}$. We use
semi-empirical tight-binding formalism to calculate electronic
states of the system for given ${\bm k}_{\|}$ and $E$ and further to
compute transmission coefficients.

\subsection{Tight-binding model}

First, we describe the construction of the tight-binding Hamiltonian
matrix for 'normal' GaAs and AlAs as well as ferromagnetic (Ga,Mn)As
layers of the heterostructure. To describe the band structure of the
bulk GaAs and bulk AlAs, we use the nearest neighbor (NN)
$sp^3d^5s^*$ tight-binding Hamiltonian (resulting in 20
spin-orbitals for each anion or cation), with the spin-orbit
coupling included.\cite{bib-9-jancu} This model reproduces correctly
the effective masses and the band structure of GaAs and AlAs in the
whole Brillouin zone. With the tight-binding Hamiltonian introduced
above, each double layer (cation + anion) is represented by
40$\times$40 matrix. It should be pointed out that the $d$ orbitals
used in our $sp^3d^5s^*$ parametrization are not related to the 3d
semi-core states and are of no use for description of Mn ions
incorporated into GaAs. The presence of Mn ions in (Ga,Mn)As is
taken into account by including the $sp$-$d$ exchange interactions
within the virtual-crystal and mean-field approximations.  In the
spirit of the tight-binding method, the effects of an external
interaction are included in the on-site diagonal matrix elements of
the tight-binding Hamiltonian. Here, the shifts of on-site energies
caused by the $sp$-$d$ exchange interaction are parameterized in
such a way that they reproduce experimentally obtained spin
splitting: $N_0\alpha=0.2$~eV of the conduction band and
$N_0\beta=-1.2$~eV of the valence band.\cite{okabayashi} It should
be, however, mentioned that since we neglect exchange interactions
between the holes, the valence spin splitting is presumably
underestimated by about 20\%.\cite{bib-7-dietl-01} The other
parameters of the model for the (Ga,Mn)As material and for the NN
interactions between GaAs and (Ga,Mn)As are taken to be the same as
for GaAs. This is well motivated because the valence-band structure
of (Ga,Mn)As with small fraction of Mn has been shown to be quite
similar to that of GaAs.\cite{okabayashi} Consequently, the valence
band offset between (Ga,Mn)As and GaAs originates only from the spin
splitting of the bands in (Ga,Mn)As. The Fermi energy in the
constituent materials is determined by the assumed carrier
concentration and is calculated from the density of states obtained
for tight-binding Hamiltonian. Our calculations of the Fermi energy
for various hole concentrations are consistent with the
corresponding results presented in Ref.~\onlinecite{bib-7-dietl-01}.
Having determined the Hamiltonian of the system, we are now in the
position to define the current and current spin polarization in the
presence of spin-orbit coupling.

\subsection{Current and current spin polarization}

For a given energy $E$ and in-plane wave-vector $k_{\|}$, the Bloch
states in the left $L$ and right $R$ leads ($i$ and $j$,
respectively) are characterized by the wave vector component
$k_{\perp}$ perpendicular to the layers and are denoted by
$|L,k_{L,\perp,i} \rangle$ and $|R,k_{R,\perp,j} \rangle$,
respectively. The indices $i$ and $j$ indicate all possible pairs
for 40 bands described by the tight-binding Hamiltonian. The
transmission probability $T_{L,k_{L,\perp,i} \to R,k_{R,\perp,j}}$
is a function of the transmission amplitude $t_{L,k_{L,\perp,i} \to
R,k_{R,\perp,j}}(E,{\bm k}_{\|})$ and group velocities in the left
and right lead, $v_{L,\perp,i}$ and $v_{R,\perp,j}$
\begin{align}
&T_{L,k_{L,\perp,i} \to R,k_{R,\perp,j}}(E,{\bm k}_{\|}) = \\
\nonumber &= \left| t_{L,k_{L,\perp,i} \to R,k_{R,\perp,j}}(E,{\bm
k}_{\|}) \right|^2 \frac{v_{R,\perp,j}}{v_{L,\perp,i}}.
\end{align}

The current flowing in the right direction can now be written as
\cite{bib-aldo-linear-response}

\begin{align}
 j_{L\to R} &=
 \frac{-e}{(2\pi)^3 \hbar}
 \int_{BZ} d^2k_{\|} dE   f_L(E)\\ \nonumber
&\!\!\!\!\!\! \sum_{\substack{ k_{L,\bot,i},k_{R,\bot,j} \\
v_{L,\bot,i},v_{R,\bot,j}>0}} T_{L,k_{L,\perp,i} \to
R,k_{R,\perp,j}}(E,{\bm k}_{\|}),
\end{align}
where $f_L$ or respectively $f_R$ are the electron Fermi
distributions in the left and right interface and $i,j$ number the
corresponding Bloch states. Plugging in the expresion given in Eq. 1
and using the time reversal symmetry

\begin{align}
T_{L,k_{L,\perp,i} \to R,k_{R,\perp,j}}&(E,{\bm k}_{\|}) = \\
\nonumber &= T_{L,-k_{R,\perp,j} \to R,-k_{L,\perp,i}}(E,{\bm
k}_{\|})
\end{align}
we get
\begin{align}
  \label{equation-lb}
  j = &\frac{-e}{(2\pi)^3 \hbar} \int_{BZ} d^2k_{\|}dE \left[f_L(E) - f_R(E)\right] \\ \nonumber
  &\sum_{\substack{ k_{L,\bot,i},k_{R,\bot,j} \\ v_{L,\bot,i},v_{R,\bot,j}>0}}
  \!\!\!\!
  \left|
    t_{L,k_{L,\bot,i} \to R,k_{R,\bot,j}}(E, {\bm k}_{\|})
  \right|^2 \frac{v_{R,\bot,j}}{v_{L,\bot,i}}.
\end{align}

Let us define the spin polarization of the outgoing Bloch state, in
respect to magnetization direction in the source lead
\begin{equation}
P_{R,k_{R,\perp,i}}(E,{\bm k}_{\|}) = \langle R,k_{R,\perp,i} |
\vec{\Omega} \cdot \vec{s} | R,k_{R,\perp,i} \rangle,
\end{equation}
where $\vec{\Omega}$ is the magnetization direction vector and
$\vec{s}$ is the spin operator. Then, we can define the spin
polarized current
\begin{align}
j_s = &\frac{-e}{(2\pi)^3 \hbar}\int_{BZ} d^2k_{\|}dE   \left[f_L(E) - f_R(E)\right]\\ \nonumber
  &\sum_{\substack{ k_{L,\bot,i},k_{R,\bot,j} \\ v_{L,\bot,i},v_{R,\bot,j}>0}}
    T_{L,k_{L,\bot,i} \to R,k_{R,\bot,j}}(E, {\bm k}_{\|}) P_{R,k_{R,\perp,i}}.
\end{align}
The  spin polarization of the coherently transmitted current is now
equal to
\begin{equation}
P_s = \frac{j_s}{j}.
\end{equation}

To calculate the current one has to determine the transmission
probability, thus the transmission amplitude $t_{L,k_{L,\bot,i} \to
k_{R,R,\bot,j}}(E, {\bm k}_{\|})$ and the group velocities
$v_{L,\bot,j}$ of the ingoing and $v_{R,\bot,j}$ of outgoing states.
These can be obtained by solving the Schr\"odinger equation for the
structure with the appropriate scattering boundary conditions. In
our studies, we follow closely the procedure detailed in
Refs.~\onlinecite{Christian} and ~\onlinecite{Aldo_review}, which we
have generalized to the case with spin-orbit coupling.

\section{Spin-dependent Zener tunneling}
\subsection{Effect of carrier densities}
The rather high $\approx 80\%$ spin polarization of the tunneling
current, obtained recently in Zener-Esaki
diodes\cite{bib-4-van-dorpe,Kohd06} opens new perspectives for
applications of electron spin injection. The degree of current spin
polarization decreases sharply with the
bias,\cite{bib-4-van-dorpe,Kohd06} an effect explained
quantitatively by our model.\cite{van-dorpe} On the other hand, it
is well known that magnetic characteristics of (Ga,Mn)As depend
strongly on both hole and manganese concentrations and that
(Ga,Mn)As films exhibit a variety of anisotropic
properties.\cite{bib-7-dietl-01,gamnas-anisotropy-1,gamnas-anisotropy-2}
It is thus obvious that the degree of spin polarization of the
tunneling current may depend on these intrinsic features of
(Ga,Mn)As layers. Indeed, we have already shown\cite{sankowski} that
a higher content of magnetic ions $x$ in Ga$_{1-x}$Mn$_x$As results
in an increase of the spin polarization of the tunneling current. In
contrast, an opposite change was obtained when the hole
concentration was increased.

\begin{figure}[h]
\epsfig{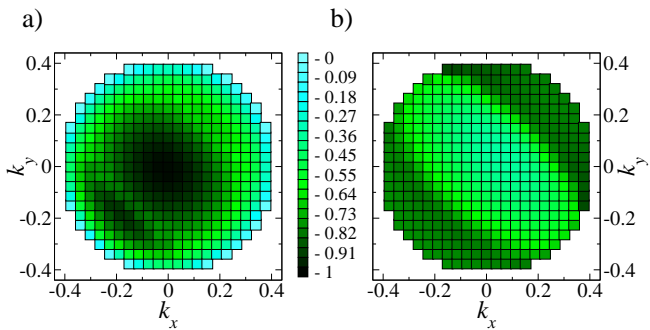} \caption{[color on-line]
Dependence of tunneling current (a) and its spin polarization (b) on
the in-plane wave vector in p-Ga$_{1-x}$Mn$_{x}$As/n-GaAs
Zener-Esaki diode in the limit of low bias. The calculation was
performed for the hole concentration  $p=3.5 \times
10^{20}$~cm$^{-3}$, electron concentration $n=10^{19}$~cm$^{-3}$,
and the saturated magnetization corresponding to the Mn content $x =
0.08$.} \label{figure-zener-inplane-wave-vector}
\end{figure}

\begin{figure}[h]
\epsfig{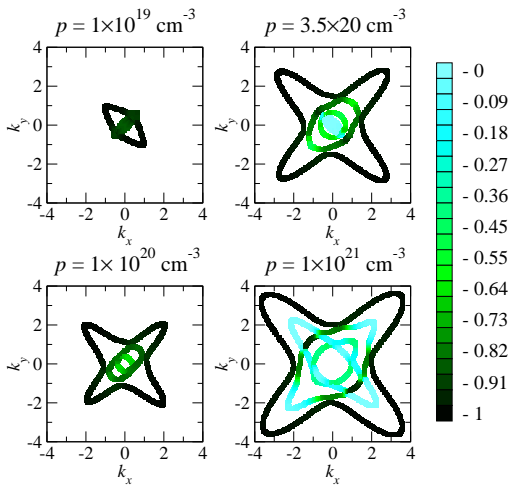}
\caption{[color on-line] Cross section of the valence bands
at the Fermi energy for various hole concentrations
$p$ in p-Ga$_{0.92}$Mn$_{0.08}$As. Color scale
denotes spin polarization. Magnetization
is taken along the $[110]$ direction.}
\label{figure-fermi-level}
\end{figure}

In order to get a better insight into processes controlling spin
polarization of the current, we have examined the dependence of
tunneling on the in-plane wave vector in the low bias limit. As
shown in Fig.~\ref{figure-zener-inplane-wave-vector}(a), the total
current is dominated by the tunneling from states close to the
$\Gamma$ point. This is because in the tunneling process the
in-plane components of ${\bm k}_{\|}$ wave vectors are conserved,
{\em i.~e.}, they have to match to the small ${\bm k}_{\|}$ vectors
at the Fermi level in the conduction band of n-type GaAs ($n =
10^{19}$~cm$^{-3}$). Turning to  current spin polarization, we note
that it decreases with the hole concentration $p$ because the higher
$p$  the smaller is the spin polarization at the Fermi level in the
vicinity of the center of the Brillouin zone. This is shown in
Fig.~\ref{figure-fermi-level} where the cross sections of the Fermi
sphere for different hole concentrations in
p-Ga$_{0.92}$Mn$_{0.08}$As with the saturated value of magnetization
are presented.  At the same time we find that the large ${\bm
k}_{\|}$ vectors are responsible for the spin polarization of the
current, as depicted in
Fig.~\ref{figure-zener-inplane-wave-vector}(b). This suggests that a
higher concentration of electrons in n-GaAs layer should result in
matching of the larger $k$ vectors and thus in higher spin
polarization of the current. Indeed, when the electron concentration
is increased from $n = 10^{19}$~cm$^{-3}$ to $n =
10^{20}$~cm$^{-3}$,  the current spin polarization becomes higher by
about 8\%.

\subsection{Anisotropic Zener tunneling - in-plane magnetization}
In this and next subsection we asses the importance of a new
mechanism called tunneling anisotropic magnetoresistance (TAMR).
This effect consists of a change in the tunnel resistance upon the
rotation of magnetization. The phenomenon, recently discovered in
structures with a single (Ga,Mn)As ferromagnetic
layer,\cite{gould,gryglas} results from the fact that tunneling
resistance depends on the relative orientation of magnetization in
respect to the direction of current and crystallographic axes. This
is due to the strong spin-orbit coupling and the highly anisotropic
Fermi surface in (Ga,Mn)As (compare Fig.~\ref{figure-fermi-level})
-- it is why TAMR was not reported for structures based on
ferromagnetic metals, where typically spin-orbit characteristic
energies are smaller than the Fermi energy.

\begin{figure}[h]
\epsfig{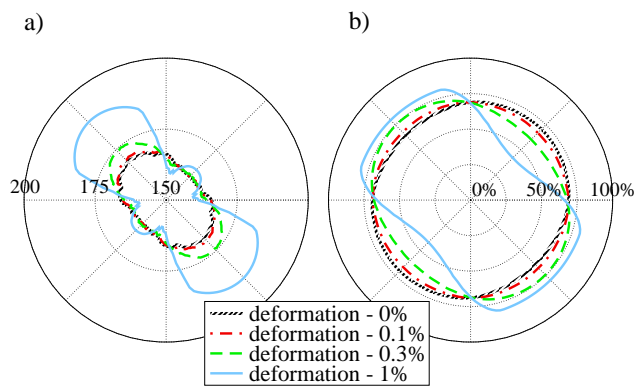} \caption{[color
on-line] Dependence of Zener tunneling current (a) and its spin
polarization (b) on the direction of in-plane magnetization without
strain and for deformations up to 1\% applied along the [110] axis.
Calculations performed for Mn content $x = 0.08$, hole concentration
$p=3.5 \times 10^{20}$~cm$^{-3}$, and electron concentration
$n=10^{19}$~cm$^{-3}$.} \label{figure-zener-deformation}
\end{figure}

We consider a simple junction consisting of several layers of p-type
Ga$_{1-x}$Mn$_x$As, $x = 0.08$, $p = 3.5 \times 10^{20}$~cm$^{-3}$,
followed by several layers of n-type GaAs, $n = 10^{19}$~cm$^{-3}$
in the weak bias limit. Interestingly, as shown in
Fig.~\ref{figure-zener-deformation} the model reveals that the
current magnitude and its spin polarization differ for magnetization
along [110] and along $[\overline{1}10]$ crystallographic axis, even
in the absence of any extrinsic deformation. This reflects the
asymmetry of the (Ga,Mn)As/GaAs interface, at which the $[110]$ and
$[\overline{1}10]$ directions are not equivalent, so that the
$T_{d}$ symmetry of the zinc-blende crystal is reduced to $C_{2v}$
for the heterostructure in question. Actually, a 6\% difference in
spin polarization of the current for the unstrained structure, which
is visible in Fig.~\ref{figure-zener-deformation}(b), agrees with
that observed experimentally.\cite{van-dorpe}

\begin{figure}[h]
\epsfig{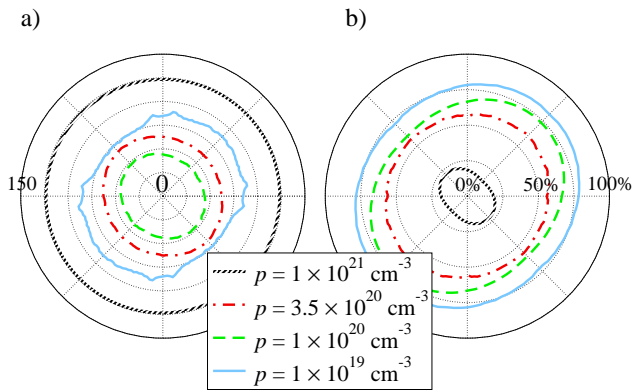} \caption{[color
on-line] Dependence of Zener tunneling current (a) and  its spin
polarization (b) on the direction of in-plane magnetization for
various hole concentrations;  Mn content $x = 0.08$ and electron
concentration $n=10^{19}$~cm$^{-3}$.}
\label{figure-zener-hole-concentration}
\end{figure}

The intrinsic anisotropy of the Zener tunneling current depends on
the hole concentration, as shown in
Fig.~\ref{figure-zener-hole-concentration}. The change in the
tunneling current upon rotation of the magnetization vector from
$[110]$ to $[\overline{1}10]$ direction increases with the decrease
of hole concentration, reaching 8.5\% for $p= 1 \times
10^{19}$~cm$^{-3}$. Such a low hole concentration can, in fact,
correspond to a depletion region in the p-(Ga,Mn)As/n-GaAs
junction,\cite{ruster,giddings} though hole localization may render
our theory invalid in this low hole concentration range. Thus, our
model predicts for (Ga,Mn)As/GaAs Zener diodes an in-plane TAMR
magnitude of the order of several percents without assuming any
extrinsic strain. Although the obtained TAMR value agrees with the
observation reported in Ref.~\onlinecite{gould}, it should be
emphasized that in that experiment (Ga,Mn)As/AlO$_x$/Au tunnel
junction was examined. Moreover, the symmetry of the experimental
TAMR effect implies the existence of an extrinsic deformation
breaking the equivalence of [100] and [010] crystallographic axes.

Typically (Ga,Mn)As films exhibit uniaxial anisotropy, whose
character implies the presence of an extrinsic trigonal distortion
along the $[110]$ axis.\cite{gamnas-anisotropy-2} A strain as small
as $0.05\%$ was found to explain the magnitude of the corresponding
uniaxial in-plane anisotropy field.\cite{gamnas-anisotropy-2} The
effect of the trigonal strain on the Zener current and its spin
polarization is presented in Fig.~\ref{figure-zener-deformation}
together with previously discussed results for unstrained
structures. As seen, the strain causes an additional in-plane
anisotropy. However, a rather strong deformation is needed to obtain
a significant dependence of spin current polarization on the
direction of the magnetization vector. Even for $0.1\%$ deformation,
which is two times larger than that evaluated in
Ref.~\onlinecite{gamnas-anisotropy-2}, the anisotropy of the spin
polarization of the current is still very small. When a strong,
$1\%$, deformation is assumed, the calculation predicts a 10\%
increase of current spin polarization for magnetization along
$[\overline{1}10]$ axis. At the same time, the obtained spin
polarization for the magnetization along $[110]$ direction is
smaller by about 30\%.

\subsection{Anisotropic Zener tunneling - out-of-plane magnetization}
The spin dependent interband tunneling is sensitive not only to the
in-plane magnetization direction. It has been shown in
Ref.~\onlinecite{gryglas} that rotation of magnetization by applying
an out-of-plane magnetic field leads also to a TAMR signal in a
Zener-Esaki diode. The magnitude of perpendicular tunneling
anisotropic magnetoresistance is defined as
\begin{equation}
\label{equation-tamr}
TAMR_{\perp} = \frac{R(H_{\perp}) - R(0)}{R(0)},
\end{equation}
where R(H$_{\perp}$) and R(0)  are the resistances for two mutually
perpendicular, out-of-plane and in-plane configurations of saturated
magnetization, {\em i.~e.}, for magnetization along $[001]$ and
$[100]$ crystallographic axis, respectively. It is worth noting that
under the presence of spin-orbit interaction, a relatively large
change in resistance is expected when the direction of magnetization
alternates from perpendicular to parallel in respect to the current,
even if the effect of epitaxial strain, which makes the $[100]$ and
$[001]$ directions non-equivalent, is disregarded. In the Boltzmann
conductance regime, the effect is known as anisotropic
magnetoresistance (AMR), and has already been studied in (Ga,Mn)As
experimentally and theoretically.\cite{Baxt02}

\begin{figure}[h]
\vspace{0.5cm} \epsfig{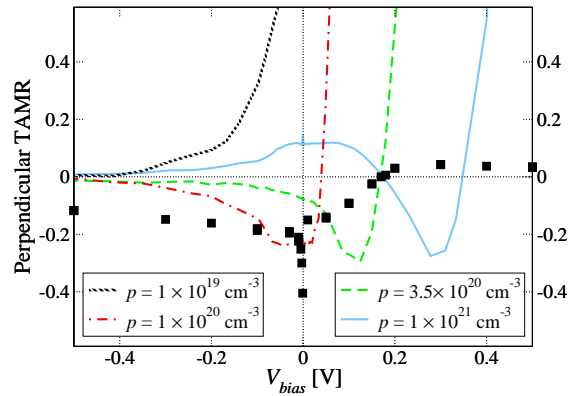}
\caption{[color on-line] The bias dependence of the relative change
in the tunneling resistance in Zener-Esaki
p-Ga$_{0.94}$Mn$_{0.06}$As/n-GaAs diode when magnetization is
rotated out-of-the-plane, for various hole concentrations;
$n=10^{19}$~cm$^{-3}$. The black squares are experimental points
from Ref.~\onlinecite{gryglas}.} \label{figure-perpendicular-tamr}
\end{figure}

The calculated relative changes of the structure resistance for the
magnetization vector flipping between perpendicular-to-the-plane and
in-plane directions are shown in
Fig.~\ref{figure-perpendicular-tamr} as a function of bias for
various hole concentrations in Ga$_{0.94}$Mn$_{0.06}$As. For
$p=3.5\times 10^{20}$~cm$^{-3}$, the maximum of the computed TAMR
effect, exceeding 20\%, is seen at small bias voltages. These
results are compared with the  experimental findings of
Ref.~\onlinecite{gryglas}, where the structure containing
ferromagnetic Ga$_{0.94}$Mn$_{0.06}$As with $T_C \approx 70$~K was
studied. According to the p-d Zener model,\cite{bib-7-dietl-01} such
a value of $T_C$ corresponds to $p\approx 10^{20}$~cm$^{-3}$.

We see in Fig.~\ref{figure-perpendicular-tamr} that the theory
describes correctly the experimental magnitude of TAMR$_{\perp}$ for
small bias at both polarizations. We see also that the computed
TAMR$_{\perp}$ tends to vanish with the increase of the reverse
bias, whereas when the forward bias is assumed, TAMR$_{\perp}$
changes sign and tends to infinity. Such change of sign for the
forward bias is also revealed experimentally, but the measured
TAMR$_{\perp}$ appears to vanish for higher values of positive bias.
This inconsistency can be explained by recalling that the computed
tunneling current stops to flow above the tunneling cutoff voltage,
which is determined by a sum of the energy distance from the hole
Fermi level $E_F^v$ to the top of the valence band in (Ga,Mn)As and
the energy difference between the bottom of the conduction band and
electron Fermi level $E_F^c$ in GaAs. In the experiment, however,
some current related to band-gap states appears to dominate near the
cutoff voltage. In turn, series bulk resistances, which are not
taken into account in the calculations, may dominate at high reverse
bias. Accordingly, standard AMR appears to contribute to the
experimental value of TAMR$_{\perp}$ in this bias
regime.\cite{gryglas}

Figure~\ref{figure-perpendicular-tamr} shows also the
TAMR$_{\perp}$ calculated for different hole concentrations in the
magnetic layer. Due to the strong p-d exchange and large spin
splitting in the (Ga,Mn)As valence band, for the low value $p=1 \times
10^{19}$~cm$^{-3}$ all spin subbands above the Fermi energy have the
same  spin polarization and thus  TAMR$_{\perp}$ does not change the sign
upon applying the positive bias. When, however, the Fermi level is
very deep in the band, in the case of large $p=1 \times
10^{21}$~cm$^{-3}$, different spin subbands contribute to the
current for various voltages and TAMR$_{\perp}$ as a function
of bias changes the sign twice.

\section{Tunneling magnetoresistance}
\subsection{Bias dependence}
\begin{figure}[h]
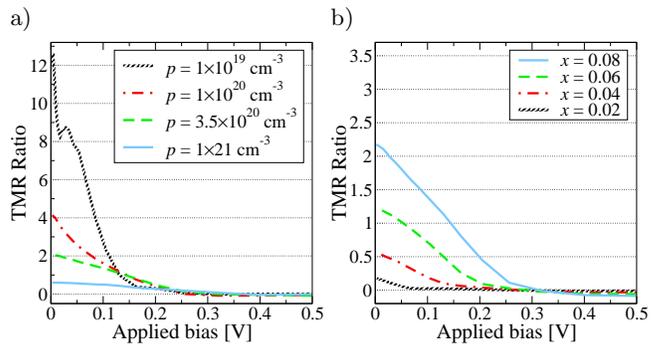

\begin{tabular}{lll}
a) & b) \vspace{0.1cm} \\
\begin{minipage}{0.23\textwidth}
\epsfig{file = gaga_0.08_all.eps, scale = 0.23}
\end{minipage}
&
\begin{minipage}{0.23\textwidth}
\epsfig{file = gaga_conc.eps, scale = 0.23}
\end{minipage}
\end{tabular}
\caption{[color on-line] The calculated bias dependence of the TMR effect in
p-Ga$_{1-x}$Mn$_{x}$As/(GaAs)$_4$/p-Ga$_{1-x}$Mn$_{x}$As trilayer
(with the GaAs barrier width $d=4$ monolayers)  for various hole
concentrations $p$ and Mn ions content $x=0.08$ (a) and for
various values of $x$ and $p=3.5 \times 10^{20}$~cm$^{-3}$ (b).}
\label{figure-tmr-bias-dependence-1}
\end{figure}

In the previous section, we have used the model to consider a device
with just one interface between magnetic (Ga,Mn)As and nonmagnetic
GaAs. The typical tunneling magnetoresistance (TMR) devices are,
however, more complicated -- they consist of a trilayer structure
with two such interfaces, for instance, two magnetic p-type
Ga$_x$Mn$_{1-x}$As contacts separated by a nonmagnetic GaAs barrier.
In such structures a strong TMR effect, {\em i.~e.}, a large
difference in the resistance of the device for two configurations:
parallel (ferromagnetic -- FM) and the antiparallel
(antiferromagnetic -- AFM) alignments of magnetizations in the
contacts, has been observed.\cite{tanaka,chiba,elsen} The TMR value
is usually described by the ratio,
\begin{equation}
\mbox{TMR} = \frac{R_{\mathrm{FM}} -
R_{\mathrm{AFM}}}{R_{\mathrm{AFM}}},
\end{equation}
where $R_{FM}$ and $R_{AFM}$ are the structure resistances for the FM
and AFM configuration, respectively. Similarly to the spin
polarization of the tunneling current in Zener-Esaki diode, TMR
increases with the content of the magnetic ions and decreases with
the concentration of the holes in (Ga,Mn)As layers.\cite{sankowski}
Unfortunately, in all experiments the observed TMR shows a rapid and
hitherto unexplained decay with the increase of the applied bias. As
shown in Fig.~\ref{figure-tmr-bias-dependence-1}, our calculations
reproduce such decay. This suggests that the dependence of TMR on
the applied bias results predominantly from the band structure
effects in this case.

\begin{figure}[h]
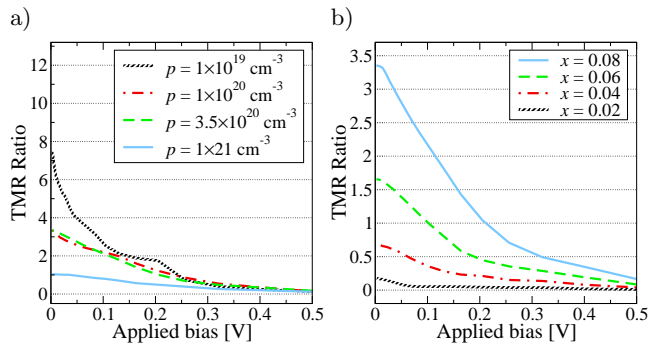

\begin{tabular}{lll}
a) & b) \vspace{0.1cm} \\
\begin{minipage}{0.23\textwidth}
\epsfig{file = gaal_0.08_all.eps, scale = 0.23}
\end{minipage}
&
\begin{minipage}{0.23\textwidth}
\epsfig{file = gaal_conc.eps, scale = 0.23}
\end{minipage}
\end{tabular}
\caption{[color on-line] The bias dependence of the TMR ratio for
p-Ga$_{1-x}$Mn$_{x}$As/(AlAs)$_4$/p-Ga$_{1-x}$Mn$_{x}$As with
$x=0.08$ and various hole concentrations (a); for various values of
$x$ and $p=3.5 \times 10^{20}$~cm$^{-3}$ (b).}
\label{figure-tmr-bias-dependence-2}
\end{figure}

In Fig.~\ref{figure-tmr-bias-dependence-1}(a) one should note a
strong dependence of TMR and its decay with applied voltage on the
hole concentration in the magnetic contacts. However, one can also
see that the hole concentration does not influence very strongly the
bias where the TMR reaches zero. This voltage, ca $0.3$~V,
corresponds to the valence band offset between
Ga$_{0.92}$Mn$_{0.08}$As and GaAs, which is determined by the spin
splitting in the valence band of the former. The presented in
Fig.~\ref{figure-tmr-bias-dependence-1}(b) TMR ratios for the
magnetic contacts with various Mn content, {\em i~e.}, with
different spin splitting, confirms this conclusion. These results
suggest that the TMR and its decrease with the applied bias can be
controlled by appropriate engineering of the band offsets in the
heterostructure, in particular, by a proper choice of the
nonmagnetic barrier. To check this prediction theoretically, we
replace in the calculations the GaAs by AlAs, which produces a
higher by $0.55$~V barrier for the holes. The results presented in
Fig.~\ref{figure-tmr-bias-dependence-2} show that, indeed, for
tunneling through the AlAs barrier, the TMR magnitude decreases with
the bias much slower than in the case of GaAs. Moreover, for a
higher tunneling barrier one can expect also higher TMR ratios.

\begin{figure}[h]
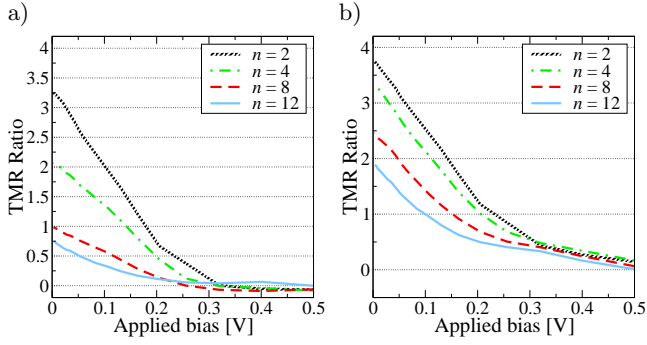

\begin{tabular}{lll}
a) && b) \vspace{0.1cm} \\
\begin{minipage}{0.23\textwidth}
\epsfig{file = gaga_0.08_width.eps, scale = 0.23}
\end{minipage}
&&
\begin{minipage}{0.23\textwidth}
\epsfig{file = gaal_0.08_width.eps, scale = 0.23}
\end{minipage}
\end{tabular}
\caption{[color on-line] The bias dependence of the TMR ratio for various
thicknesses $d$ of the barrier layer in (a)
p-Ga$_{0.92}$Mn$_{0.08}$As/(GaAs)$_d$/p-Ga$_{0.92}$Mn$_{0.08}$As;
(b)
p-Ga$_{0.92}$Mn$_{0.08}$As/(AlAs)$_d$/p-Ga$_{0.92}$Mn$_{0.08}$As}
\label{tmr-width}
\end{figure}

A related behavior can be seen in the study of the dependence of the
TMR effect on the width of the barrier, presented in
Fig.~\ref{tmr-width}. In agreement with experimental observations,\cite{tanaka}
the calculated TMR drops rapidly when the barrier becomes wider.
However, for AlAs which forms a higher barrier,
the decrease of the TMR ratio with the number of barrier
monolayers is much weaker.

\subsection{Anisotropy of tunneling magnetoresistance}
\begin{figure}[h]
\epsfig{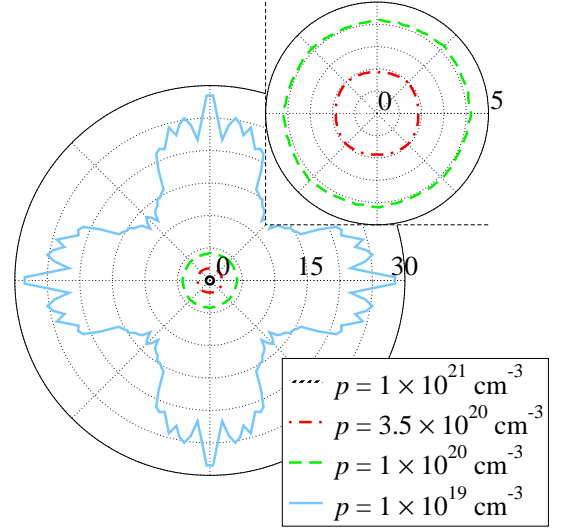} \caption{[color on-line] Dependence of
TMR ratio on the direction of in-plane magnetization
for various hole concentrations in p-Ga$_{0.92}$Mn$_{0.08}$As/(GaAs)$_4$/p-Ga$_{0.92}$Mn$_{0.08}$As
trilayer structure.} \label{figure-tmr-concentration}
\end{figure}

The calculated dependence of TMR on the in-plane direction of the
magnetization vector is shown in
Fig.~\ref{figure-tmr-concentration}. It is seen that in the case of
TMR the [100] and [110] magnetization directions remains not
equivalent, while TMR is identical for $[110]$ and
$[\overline{1}10]$. Thus, the $D_{2d}$ symmetry is recovered if two
interfaces are involved, in contrast to  the case of spin current
polarization in the Esaki-Zener diode, where $C_{2v}$ symmetry of a
single zinc-blende interface led to the non-equivalence of the
$[110]$ and $[\overline{1}10]$ directions, as discussed in the
previous section.

\begin{figure}[h]
\epsfig{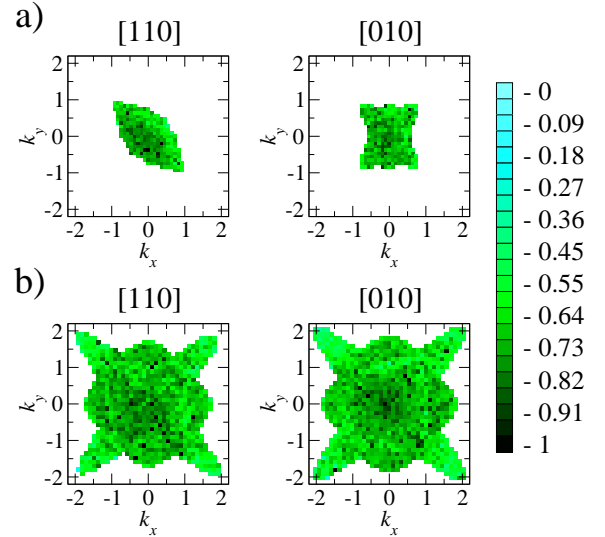} \caption{[color on-line]
Dependence of tunneling current on the direction of the in-plane
wave vector for antiparallel configuration of magnetizations (AFM)
along the $[110]$ and $[010]$ crystallographic directions, as
indicated in the plots, and for hole concentrations $1 \times
10^{19}$~cm$^{-3}$ (a) and $3.5 \times 10^{20}$~cm$^{-3}$ (b).}
\label{figure-tmr-wave-vectors}
\end{figure}

As shown in Fig.~\ref{figure-tmr-concentration}, the in-plane
anisotropy of TMR depends crucially on the hole concentration in the
magnetic layer. For hole concentrations $p$ in the range of
$~10^{20}$~cm$^{-3}$ the obtained anisotropy of TMR is below 10\%,
however, for low concentrations, $p = 10^{19}$~cm$^{-3}$, it becomes
as strong as 250\%. The reason for this behavior becomes clear when
we look at Fig.~\ref{figure-tmr-wave-vectors} that shows the
dependence of the tunneling current in the AFM configuration on the
in-plane wave vector for different directions of magnetization. In
Fig.~\ref{figure-tmr-wave-vectors}(a), {\em i. e.}, in the case of
$p=10^{19}$~cm$^{-3}$ the region of the Brillouin zone that takes
part in the tunneling is strongly dependent on the magnetization
direction, in contrast to the results for higher hole concentrations
presented in Fig.~\ref{figure-tmr-wave-vectors}(b). It should be
stressed that such behavior has been obtained only for the AFM
alignment. The calculated tunneling current in the FM configuration
does not virtually depend on the direction of magnetization -- only
a very small difference between $[110]$ and $[010]$ direction has
been found.

\begin{figure}[h]
\epsfig{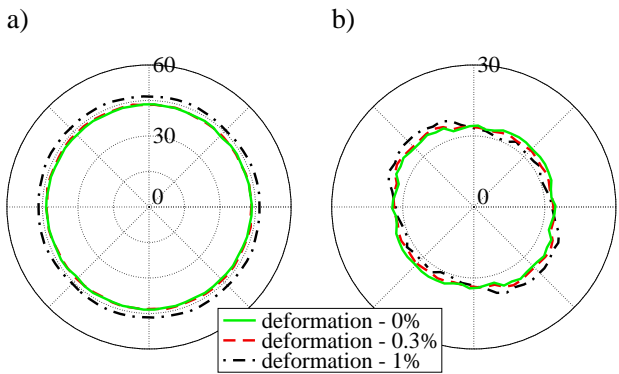}
\caption{[color on-line] Dependence of tunneling current on the
direction of in-plane magnetization for (a) parallel (FM) and (b)
antiparallel (AFM) magnetization orientations in
p-Ga$_{0.92}$Mn$_{0.08}$As/(GaAs)$_4$/p-Ga$_{0.92}$Mn$_{0.08}$As
trilayer structures trigonally distorted along the [110]
crystallographic axis; $p=3.5 \times 10^{20}$~cm$^{-3}$.}
\label{figure-tmr-deformations}
\end{figure}

A similar effect can be also noticed in the calculated dependence of
the TMR ratio on strain, as shown in
Fig.~\ref{figure-tmr-deformations}. Here again we see that upon
trigonal strain the tunneling current becomes anisotropic only for
the AFM alignment of magnetization in the two magnetic contacts.

\subsection{Tunneling anisotropic magnetoresistence}

\begin{figure}[h]
\vspace{0.4cm} \epsfig{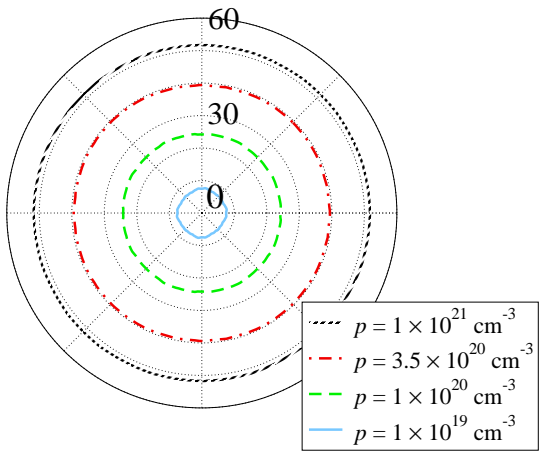}
\caption{[color on-line] Dependence of tunneling current on the
direction of in-plane magnetization structures with parallel
magnetization configuration (FM) for various hole concentrations in
p-Ga$_{0.92}$Mn$_{0.08}$As/(GaAs)$_4$/p-Ga$_{0.92}$Mn$_{0.08}$As
trilayer structure; $p=3.5 \times 10^{20}$~cm$^{-3}$.}
\label{figure-tmr-tamr}
\end{figure}

The results presented above show that the computed anisotropy of TMR
results exclusively from the anisotropy of tunneling in the AFM
configuration. Thus, this anisotropy cannot explain the in-plane
TAMR effect observed in the (Ga,Mn)As/GaAs/Ga,Mn)As
structures,\cite{ruster,giddings} {\em i.~e.}, the difference in the
resistance of the structures in FM configuration between the $x$ and
$y$ directions. As stated before, we  obtain only a very small
difference in the FM tunneling current between the $[110]$ and
$[010]$ direction. Although this anisotropy can be increased to
about 15\% by assuming a low hole concentration of $p =
1\times10^{19}$~cm$^{-3}$ (compare Fig.~\ref{figure-tmr-tamr}), the
depletion effects would not affect the anisotropy directions
determined by the symmetry of the structure.

\begin{figure}[h]
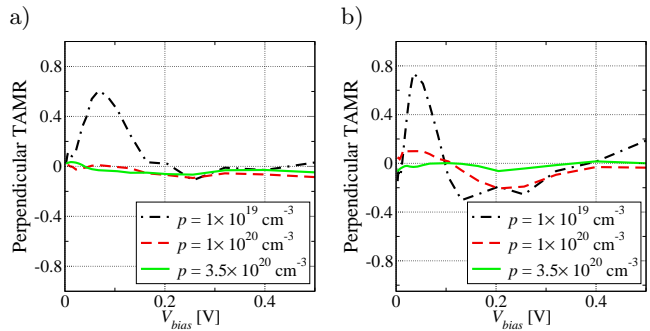

\begin{tabular}{lll}
a) && b) \vspace{0.1cm} \\
\begin{minipage}{0.23\textwidth}
\epsfig{file = gaga_tmr_tamr.eps, scale = 0.22}
\end{minipage}
&&
\begin{minipage}{0.23\textwidth}
\epsfig{file = gaal_tmr_tamr.eps, scale = 0.22}
\end{minipage}
\end{tabular}
\caption{[color on-line] Bias dependence of the TAMR$_{\perp}$ ratio for TMR
structure consisting of two magnetic p-Ga$_{0.92}$Mn$_{0.08}$As
layers separated by (a) GaAs and (b) AlAs barrier layer; $p=3.5 \times 10^{20}$~cm$^{-3}$.}
\label{figure-tmr-tamr-out-of-plane}
\end{figure}

However, the calculated tunneling current for magnetization vector
perpendicular to the plane differs from the current calculated for
the in-plane magnetization vector even for the FM configuration.
Using the TAMR$_{\perp}$ ratio defined in Eq.~\eqref{equation-tamr},
we have calculated the perpendicular TAMR for the
(Ga,Mn)As/GaAs/(Ga,Mn)As and (Ga,Mn)As/AlAs/(Ga,Mn)As structures. The
results are presented in Fig.~\ref{figure-tmr-tamr-out-of-plane}. In
the case of the GaAs spacer, for hole concentrations of about
$1\times 10^{20}$~cm$^{-3}$ a very small effect that weakly depends
on the applied bias can be observed. Still, for small hole
concentration we see a positive TAMR$_{\perp}$ of the order of 60\%.
The difference between the out-of-plane and in-plane resistances of
the same sign and of about 12\% was experimentally observed in
Ref.~\onlinecite{giddings}. As shown in
Fig.~\ref{figure-tmr-tamr-out-of-plane}, again a higher barrier,
{\em i.~e.}, the AlAs spacer, should enhance the TAMR$_{\perp}$
effect.

\section{Summary}

We have developed the model of quantum transport in spatially
modulated structures of hole-controlled diluted ferromagnetic
semiconductors, taking into account relevant features of the band
structure within the tight-binding approximation. The model
disregards disorder and effects of carrier--carrier interactions, so
that it is applicable to the carrier density range and length
scales, where localization effects are unimportant. The computation
results presented in this and our previous
papers\cite{van-dorpe,sankowski} demonstrate that many of
experimentally important effects, such as large magnitudes of both
spin polarization of the tunneling current in Zener-Esaki diodes and
TMR ratio in trilayer structures can be understood within the
proposed model. Furthermore, the theory describes quantitatively a
fast decay of the spin polarization of the current and TMR with the
bias voltage without invoking inelastic processes. However, these
processes together with heating of Mn spin subsystem may become
crucial in the highest bias regime. The detail studies of anisotropy
effects reveal the presence of $C_{2v}$ symmetry in the magnitude of
current spin polarization in the Zener-Esaki diode. This indicates a
noticeable importance of inversion asymmetry terms specific to
interfaces and zinc-blende structure in tunneling structures. These
effects are not taken into account within the standard $kp$-type
approaches. According to our findings, if strain is not excessively
large,  the dominant anisotropy appears when the direction of
magnetization changes from parallel to perpendicular to the current,
in a full analogy to AMR. Finally, we have used the model to predict
theoretical conditions for improving the performance of the studied
devices. Our results indicate that an attempt to increase the
electron concentration in the n-GaAs layer of the Zener-Esaki tunnel
junction should pay off in an increase of the spin polarization of
the current. For the trilayers, the calculations suggest that
reducing barrier thickness and increasing barrier height may result
in higher values of TMR and its slower decay with the applied bias.

\section*{Acknowledgments}
We thank Hideo Ohno, Fumihiro Matsukura, Wim Van Roy, and Albert Fert for valuable
discussions. This work was partly supported by the
EC project NANOSPIN (FP6-2002-IST-015728). Computations were carried
out exploiting resources and software of Interdisciplinary Center of
Mathematical and Computer Modelling (ICM) in Warsaw.

\end{document}